\documentclass{article}
\usepackage{dcase2021_techrep,amsmath,graphicx,url,times,booktabs, tabularx}
\usepackage{enumitem}
\usepackage{float}
\usepackage{multirow}
\usepackage{booktabs}
\usepackage{xcolor}
\usepackage{color, colortbl}

\definecolor{Gray}{gray}{0.9}

\title{Task 1A DCASE 2021: Acoustic Scene Classification with mismatch-devices using squeeze-excitation technique and low-complexity constraint}

\name{Javier Naranjo-Alcazar$^{1,2}$,
      Sergi Perez-Castanos$^{2}$,
      Maximo Cobos$^{2}$,
      Francesc J. Ferri$^{2}$,
      Pedro Zuccarello$^{1}$
      }

\address{$^1$ Instituto Tecnológico de Informática, València, Spain \{jnarnajo, pzuccarello\}@iti.es\\         
$^2$ Universitat de Val\`encia, Burjassot, Spain, \{pecaser@alumni.uv.es, \{maximo.cobos, francesc.ferri\}@uv.es\}\\
 }

\begin{document}

\ninept
\maketitle

\begin{sloppy}

\begin{abstract}
Acoustic scene classification (ASC) is one of the most popular problems in the field of machine listening. The objective of this problem is to classify an audio clip into one of the predefined scenes using only the audio data. This problem has considerably progressed over the years in the different editions of DCASE. It usually has several subtasks that allow to tackle this problem with different approaches. The subtask presented in this report corresponds to a ASC problem that is constrained by the complexity of the model as well as having audio recorded from different devices, known as mismatch devices (real and simulated). The work presented in this report follows the research line carried out by the team in previous years. Specifically, a system based on two steps is proposed: a two-dimensional representation of the audio using the Gamamtone filter bank and a convolutional neural network using squeeze-excitation techniques. The presented system outperforms the baseline by about 17 percentage points.
\end{abstract}


\section{Introduction}
\label{sec:intro}

\par Extracting information from audio signals can be a great improvement in existing applications or future products (home assistants, wildlife monitoring, autonomous cars, etc.). Machine listening is understood as the set of algorithms that are capable of extracting relevant information from audio. One of the most common tasks in this field is known as Acoustic Scene Classification (ASC) \cite{mesaros2018acoustic, mesaros2019acoustic, valenti2016dcase, barchiesi2015acoustic}. The ultimate goal is to extract context information from the audio, more specifically, to predict the location where the audio is produced (park, metro station, airport, etc.). This problem has been addressed in all previous editions of DCASE, and has been modified with different restrictions. In this report an ASC system is designed to be limited by the size of the model and with the extra difficulty that the audios used in the training come from different audio sources (mismatch devices).


\par The motivation of DCASE 2021 Task 1a is to create an acoustic scene classifier that should work in real-time (low-complexity consideration) and capable of using different recording sources (microphones) \cite{Martinmorato2021lowcomplexity}. This subtask can be understood as a merge of both subtask in the 2020 edition in which the mismatch problem had no restriction on the model, and on the other hand, the low complexity subtask only used audios from the same recording source.

\par The approach proposed in this work consists in a CNN implemented with squeeze-excitation modules feed with a 2D audio representation using the Gammatone filter bank. The model is converted to TFLite format in order to accomplish the model size restriction. More information on the proposed framework is presented in Section~\ref{sec:method}, while the results obtained in the development stage are presented in Section~\ref{sec:results}. Some conclusions are drawn in Section~\ref{sec:conclusion}.




\section{Method}\label{sec:method}

\subsection{Audio Representaion}\label{subsec:audio_repr}

\par Following the idea of last year submissions \cite{perezcastanos2019cnn, naranjo2020task} a Gammatone filter bank-based representation has been chosen for providing a slightly superior performance than other alternatives (e.g. Mel-scale filter banks) in preliminary tests. 

\par All representations are calculated with a window size of 40 ms with 50\% overlapping, using a sampling rate of 44.1 kHz and 64 frequency bins. All frequency bins are normalize with 0 mean and standard deviation equal to 1 using all the provided data. Gammatone representations were computed by using the Auditory Toolbox presented in \cite{slaney1998auditory} with Python implementation.


\subsection{Convolutional Neural Network}\label{subsec:SE}

\par The convolutional network trained with the audio information is composed of blocks defined as \emph{Conv-StandardPOST}. These blocks were proposed in \cite{naranjo2020acoustic}. The aim of these blocks is to achieve improved accuracy by recalibrating the internal feature maps using residual \cite{he2016deep} and squeeze-excitation techniques \cite{hu2018squeeze, roy2018concurrent}. For more insight about this choice, please see \cite{naranjo2020acoustic} where \emph{Conv-StandardPOST} is fully explained and compared to other competing blocks. The architecture of the network can be seen in Table~\ref{tab:network}

\begin{table}[]
\centering
\begin{tabular}{c }
\toprule
 Gammatone representation $(64 \times T \times 1)$ \\ \midrule
\rowcolor{Gray} \emph{Conv-StandardPost} $(\#40, 3, \rho=2)$ \\ 
 \midrule
 Max Pooling $(1, 10)$ \\ \midrule
 Dropout $(0.3)$ \\ \midrule
\rowcolor{Gray} \emph{Conv-StandardPost} $(\#40, 3, \rho=2)$ \\ 
 \midrule
 Max Pooling $(1, 10)$ \\ \midrule
 Dropout $(0.3)$ \\ \midrule
 Global Average Pooling \\ \midrule
 Classification $(10)$ \\
 \bottomrule
\end{tabular}
\caption{Network architecture. \emph{Conv-StandardPost} is denoted with the number of filters ($\#$), the kernel size and the ratio of the squeeze-excitation module ($\rho$). The Max Pooling layer is defined by the pool size and the Dropout by the rate. The classification layer corresponds to a Dense with $10$ units.}
\label{tab:network}
\end{table}

\subsection{Experimental details}\label{subsec:exp_details}

\subsubsection{Training}

The optimizer used is Adam \cite{kingma2014adam}. The loss used is the one known as Focal Loss \cite{lin2017focal}. This loss function assigns greater emphasis to those samples that are not classified correctly, forcing the system to correctly classify the more challenging samples (those related to devices with lower resolution). The hyperparameters are set as $\alpha=0.25$ and $\gamma=2$. During training, the learning rate (which starts at 0.001) is modified by a factor of 0.5 if the validation accuracy does not improve for 20 epochs. Training ends if this metric is not improved for 50 epochs. The maximum number of epochs is 500.

\subsubsection{Dataset}

The dataset provided for the task is known as TAU Urban Acoustic Scenes 2020 Mobile \cite{Mesaros2018_DCASE}. In turn, this dataset is divided into two splits, the development split and the evaluation split. While the development split contains scenes recorded in 10 cities, the evaluation one contains scenes from 12 cities (there are two cities unseen in the development set). The development split contains audio recorded from 3 real devices and 6 simulated ones. The total amount of hours present in this specific split is 64 hours. The audio is provided in mono, 44.1 kHz of sampling rate and 24-bit format.

\section{Results}\label{sec:results}

\par The results obtained by the system proposed can be seen in Table~\ref{tab:results}. The proposed approach surpass the baseline by 17 percentage points by only having 5 KB more than the baseline regarding model complexity.

\begin{table}[H]
\centering
\begin{tabular}{ccc}
\toprule
& Accuracy (\%) & Model size (KB) \\ \midrule
\rowcolor{Gray} Challenge Baseline & 47.70 & 90.30 \\ 
 \midrule
 Proposed system & \textbf{64.18} & 95.96  \\
 \bottomrule
\end{tabular}
\caption{Accuracy (\%) results obtained compare with the proposed baseline}
\label{tab:results}
\end{table}

\section{Conclusion}\label{sec:conclusion}

Understanding the sounds around us can be a great improvement in a multitude of applications. These solutions must deal with certain issues that may arise. In this task, scene classification problem is proposed with the extra issues of mismatch devices (available audios come from different sources) and complexity constraint (intended to be deployed in real-time solutions on edge devices for example).

In this an ASC system based on the Gammatone representation of the audio and a slim neural network using squeeze-excitation techniques is presented to improve its performance, which is then converted to TFLite format to reduce its size.

\section{Acknowledgements}\label{sec:acknow}

\par The participation of Dr. Cobos  and Dr. Ferri is supported by ERDF and the Spanish Ministry of Science,
Innovation and Universities under Grant RTI2018-097045-B-C21, as well as grants AICO/2020/154 and AEST/2020/012 from Generalitat Valenciana.g

\bibliographystyle{IEEEtran}
\bibliography{refs}

\end{sloppy}
\end{document}